\documentclass[12pt,preprint,letterpaper,nofootinbib,superscriptaddress, preprintnumbers]{revtex4-1}

\usepackage{graphicx} 
\usepackage{epstopdf}
\usepackage{bm}
\usepackage[colorlinks=true,linktocpage=true,linkcolor=blue,citecolor=blue,urlcolor=blue]{hyperref}
\usepackage{textcomp}
\usepackage{color}
\usepackage{xcolor}
\usepackage{tabularx}
\usepackage{tabu}
\usepackage{amsmath,amssymb,mathtools}
\usepackage{feynmp-auto}
\usepackage[normalem]{ulem}

\begin{document}

\preprint{IFT-UAM/CSIC-24-162}

\title{NLO SMEFT Electroweak Corrections to Higgs Decays to 4 Leptons in the Narrow Width Approximation}

\def\BNL{High Energy Theory Group, Physics Department, 
    Brookhaven National Laboratory, Upton, NY 11973, USA}

\def\IFT{Departamento de F\'{i}sica Te\'{o}rica and Instituto de F\'{i}sica Te\'{o}rica UAM/CSIC, Universidad Aut\'{o}noma de Madrid, Cantoblanco, 28049, Madrid, Spain}

\def\YITP{C. N. Yang Institute for Theoretical Physics, Stony Brook University, Stony Brook, NY, 11794, USA}

\author{Sally Dawson}
\email{dawson@bnl.gov}
\affiliation{\BNL}

\author{Matthew Forslund}
\email{matthew.forslund@stonybrook.edu}
\affiliation{\BNL}
\affiliation{\YITP}

\author{Pier Paolo Giardino}
\email{pier.giardino@uam.es}
\affiliation{\IFT}
\begin{abstract}
\noindent
Some of the most precise measurements of Higgs boson couplings are from the Higgs decays to 4 leptons, where deviations from the Standard Model predictions can be quantified in the framework of the Standard Model Effective Field Theory (SMEFT). In this work, we present a complete next-to-leading order (NLO) SMEFT electroweak calculation of the rate for $H\rightarrow  \ell^+\ell^- Z$ which we combine with the NLO SMEFT result for $Z\rightarrow\ell^+\ell^-$ to obtain  the  NLO rate for the $H\rightarrow$ 4 lepton process in the narrow width approximation.  The NLO calculation provides sensitivity to a wide range of SMEFT operators that do not contribute to the rate at lowest order and demonstrates the importance of including correlations between the effects of different operators when extracting limits on SMEFT parameters.  We show that the extraction of the Higgs tri-linear coupling from the decay $H\rightarrow  \ell^+ \ell^- Z, \ Z\rightarrow \ell^+\ell^-$ in the narrow width approximation strongly depends  on the contributions of other operators that first occur at NLO.
\end{abstract}

\maketitle

\section{Introduction}

Since the discovery of the Higgs boson at the LHC in 2012 there has been an intense effort, both theoretically and experimentally, to obtain precise  measurements and predictions for Higgs properties.  The mass is measured to ${\cal{O}}(.1\%)$~\cite{ATLAS:2023oaq,CMS:2024eka} and Higgs coupling measurements to $3^{rd}$ generation fermions and gauge bosons vary from ${\cal{O}}(5-20\%)$ accuracy~\cite{ATLAS:2024fkg,CMS:2022dwd} with prospects for future measurements at the HL-LHC at the few percent level~\cite{Dawson:2022zbb}.  One of the most precisely measured quantities is the branching ratio of the Higgs boson to 4 leptons, which is known to ${\cal{O}}(10\%)$~\cite{ATLAS:2020rej,CMS:2023gjz}.  At lowest order (LO), this rate is  sensitive to the $Z$ boson coupling to the Higgs boson and has been extensively used to probe anomalous Higgs -gauge boson couplings~\cite{Ellis:2020unq,Celada:2024mcf,Almeida:2021asy,Grinstein:2013vsa,Boselli:2017pef}.  Including a subset of higher order corrections, the Higgs decay to 4 leptons has  been used to indirectly probe the Higgs tri-linear coupling~\cite{Degrassi:2017ucl,Degrassi:2016wml,Maltoni:2017ims}.  The decay to 4 leptons also depends on the couplings of the leptons to the $Z$ boson, but these interactions are stringently restricted by $Z$ pole measurements~\cite{ALEPH:2005ab} and thus play a smaller role.   

In the Standard Model (SM), the rate, along with the differential distributions, for $H\rightarrow $ 4 leptons is well known to next-to-leading order in the electroweak theory~\cite{Bredenstein:2006rh} and can be straightforwardly obtained from the public code, PROPHECY4f~\cite{Denner:2019fcr}.  To  look for effects beyond  SM physics through precision measurements of Higgs decays, it is useful to employ the SM Effective Field Theory (SMEFT)~\cite{Brivio:2017vri}, where new physics effects are expressed as an expansion around the SM,
\begin{equation}
\label{eq:smeftdim6}
\mathcal{L}_\text{SMEFT}^6=\mathcal{L}_\text{SM}+\sum_{i}{C_i\over\Lambda^2}\mathcal{O}_i+...
\end{equation}
where $\mathcal{O}_i$ consists of the complete set of dimension-6 $SU(3)\times SU(2)\times U(1)$ invariant operators constructed out of SM fields, $\Lambda$ is an arbitrary scale typically taken as $1~\text{TeV}$,  $C_i$ are the unknown Wilson coefficients  that contain information about the UV structure of the theory and we neglect higher dimension operators.  Since there is no hint of new physics at the LHC, we assume that the scale $\Lambda$ is well separated from the weak scale. We note that at dimension-6, observables only depend on  the ratio, $C_i/\Lambda^2$, and so deriving a sensitivity to a new scale requires assumptions about the couplings $C_i$.

In this work, we compute  $H\rightarrow \ell^+\ell^- Z$, ($\ell=e,\mu$), in the SMEFT  at  NLO in the electroweak couplings in order to probe effects of beyond the Standard Model (BSM) physics through a precision measurement of the decay rate.  This extends the NLO electroweak SMEFT calculation of $H\rightarrow ZZ$~\cite{Dawson:2018pyl} to the relevant case for the physical $m_H$.
The leading order (LO) SMEFT rates and kinematic distributions are altered by the NLO electroweak corrections, but even more interesting is the sensitivity to new interactions that first enter at NLO SMEFT. There are 
$66$ independent CP conserving operators in the Warsaw basis 
that contribute at NLO, which potentially dilutes the sensitivity to any specific operator (such as the operator generating the Higgs tri-linear coupling). We employ the narrow width approximation to relate $H\rightarrow  \ell^+\ell^- Z$ to $H\rightarrow  \ell^+ \ell^- Z, \ Z\rightarrow \ell^+\ell^-$, using the known NLO dimension-6 results for $Z\rightarrow \ell^+\ell^-$\cite{Dawson:2019clf,Dawson:2022bxd,Bellafronte:2023amz}.

In Section~\ref{sec:smeft}, we review the dimension-6 SMEFT as used in this paper and in Section~\ref{sec:calc} we describe the NLO electroweak calculation of $H\rightarrow  \ell^+\ell^- Z$. The virtual contributions  can be obtained from the NLO electroweak calculation of $e^+e^-\rightarrow ZH$~\cite{Asteriadis:2024xts,Asteriadis:2024qim} by crossing, while the real emission contribution requires integration over the four-body final state phase space. Section~\ref{sec:results} contains numerical results in a format that can be easily implemented into Monte Carlo codes.  We also demonstrate the interplay of the narrow width results for $H\rightarrow $ 4 leptons with precision $Z$ pole limits from $Z\rightarrow e^+e^-$ and discuss the accuracy of the narrow width approximation for obtaining an NLO SMEFT result for $H\rightarrow  \ell^+ \ell^- Z, \ Z\rightarrow \ell^+\ell^-$. We emphasize the need to consistently include NLO electroweak effects in SMEFT studies and provide  an outlook of  future prospects for including NLO SMEFT results in  global fits and projections in the conclusion.

\section{SMEFT Basics}
\label{sec:smeft}
In our calculation we use the SMEFT dimension-6 Lagrangian of eq.~\eqref{eq:smeftdim6} expressed in terms of the Warsaw basis~\cite{Buchmuller:1985jz,Grzadkowski:2010es}, following the notation of~\cite{Dedes:2017zog}.  
We do not impose any flavor structure on the SMEFT operators, however we take the CKM matrix to be diagonal; this choice effectively restricts the number of flavorful operators that can appear in the calculation~\cite{Greljo:2022cah,Bellafronte:2023amz}. 
We chose to work in the $(G_\mu,m_W,m_Z)$ input scheme and the vacuum expectation value, $v_T$, is defined to be the minimum of the potential at all orders in the loop expansion.

The presence of the SMEFT operators changes the relations between the $SU(2)$ and $U(1)$ gauge couplings $g_2$ and $g_1$ entering the Lagrangian, $\alpha$, and our input parameters. The new relationships, valid to ${\cal{O}}({1 /\Lambda^2})$, are~\cite{Asteriadis:2022ras,Hays:2018zze}
\begin{align}
\label{eq:parshift}
\begin{split}
 {g_2} &= 2M_W (\sqrt{2}G_\mu)^{1/2}\biggl\{1+ \frac 12 X_H\biggr\} \, , \\
 {g_1} &= 2(\sqrt{2}G_\mu)^{1/2} \sqrt{M_Z^2-M_W^2}\biggl\{1+\frac 12 X_H\biggr\} \\
    &\hspace{10pt}-\frac{1}{2 (\sqrt{2}G_\mu )^{1/2}\Lambda^2}\biggl\{4M_W C_{\phi WB}+\frac{M_Z^2}{\sqrt{M_Z^2-M_W^2}}C_{\phi D}\biggr\} \, ,\\
    4 \pi \alpha  &= {G_\mu}\sqrt {2}
     \left(1+ X_H  \right) 4 M_W^2 \left( 1- \frac{M_W^2}{M_Z^2}\right)\\
     &-\frac{2M_W^3} {M_Z^2\Lambda^2}\biggl\{M_W C_{\phi D} +{4}\sqrt{M_Z^2-M_W^2}C_{\phi W B}\biggr\}  \, \\
     v_T^2 &=  {(1+\Delta r)\over\sqrt{2}G_\mu}(1-X_H)
\end{split}
\end{align}
where 
\begin{align}
    X_H\equiv \frac{1}{\sqrt{2}G_\mu\Lambda^2}\biggl\{
    C_{ll}[1221]- (C_{\phi l}^{(3)}[11]+C_{\phi l}^{(3)}[22])\biggr\} \, .
    \label{eq:xdef}
\end{align}
The indices $[ii]$ and $[ijji]$, etc, are flavor indices.
We note that $e=\sqrt{4\pi\alpha}$ is not an independent parameter of the model ($e$ is defined as the coupling of the electron to the photon in SMEFT).
Since the vev, $v_T$, is defined to be the minimum of the potential, the relationship between $v_T$ and $G_\mu$ receives corrections at one-loop that have SMEFT contributions along with the well known SM result.
An explicit expression for the dimension-6 one- loop SM and SMEFT results for $\Delta r$ can be found in the appendix of \cite{Dawson:2018pyl}.

\section{Calculation}

\label{sec:calc}

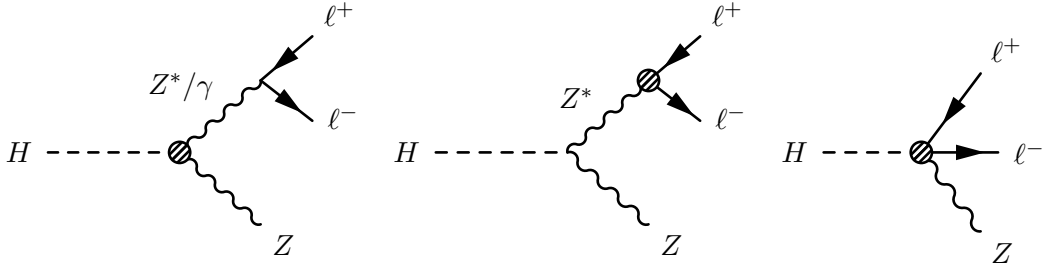
\begin{figure}
    \centering
\begin{fmffile}{lo_zgam}
  \begin{fmfgraph*}(100,120)
    \fmfstraight
    \fmfleft{i2,g,i1}
    \fmfright{o4,o3,o2,o1}
    \fmftop{top}
    \fmfbottom{bot}
    \fmf{dashes,tension=1.6}{g,h}
    \fmf{photon,tension=1.3,label=$Z^*/\gamma$,l.s=right}{v1,h}
    \fmf{photon,tension=1.3,l.s=right}{h,v2}
    \fmf{phantom,tension=1.0}{top,v1}
    \fmf{phantom,tension=1.0}{v2,bot}
    \fmfshift{16 down}{o1}
    \fmfshift{ 8 down}{o2}
    \fmfshift{ 8 up}{o3}
    \fmfshift{16 up}{o4}
    \fmf{fermion,tension=1.8}{o1,v1,o2}
    \fmf{phantom,tension=1.8}{o4,v2,o3}
    \fmfblob{0.08w}{h}
    \fmfv{l=$H$}{g}
    \fmfv{l=$\ell^+$}{o1}
    \fmfv{l=$\ell^-$}{o2}
    \fmfv{l=$Z$}{v2}
  \end{fmfgraph*}
\end{fmffile} 
\hspace{1cm} \begin{fmffile}{lo_z}
  \begin{fmfgraph*}(100,120)
    \fmfstraight
    \fmfleft{i2,g,i1}
    \fmfright{o4,o3,o2,o1}
    \fmftop{top}
    \fmfbottom{bot}
    \fmf{dashes,tension=1.6}{g,h}
    \fmf{photon,tension=1.3,label=$Z^*$,l.s=right}{v1,h}
    \fmf{photon,tension=1.3,l.s=right}{h,v2}
    \fmf{phantom,tension=1.0}{top,v1}
    \fmf{phantom,tension=1.0}{v2,bot}
    \fmfshift{16 down}{o1}
    \fmfshift{ 8 down}{o2}
    \fmfshift{ 8 up}{o3}
    \fmfshift{16 up}{o4}
    \fmf{fermion,tension=1.8}{o1,v1,o2}
    \fmf{phantom,tension=1.8}{o4,v2,o3}
    \fmfblob{0.08w}{v1}
    \fmfv{l=$H$}{g}
    \fmfv{l=$\ell^+$}{o1}
    \fmfv{l=$\ell^-$}{o2}
    \fmfv{l=$Z$}{v2}
  \end{fmfgraph*}
\end{fmffile}
\hspace{1cm}
\begin{fmffile}{lo_4pt}
  \begin{fmfgraph*}(66,120)
    \fmfleft{i0}
    \fmfright{o3,o2,o1}
    \fmftop{top}
    \fmfbottom{bot}
    \fmf{dashes,tension=2}{i0,v1}
    \fmfshift{30 down}{o1}
    \fmfshift{30 up}{o3}
    \fmf{fermion,tension=1}{o1,v1,o2}
    \fmf{photon}{v1,o3}
    \fmfblob{8}{v1}
    \fmfv{l=$H$}{i0}
    \fmfv{l=$\ell^+$}{o1}
    \fmfv{l=$\ell^-$}{o2}
    \fmfv{l=$Z$}{o3}
  \end{fmfgraph*}
\end{fmffile}
\vspace{-1cm}
\caption{Diagrams contributing at LO in the dimension-6 SMEFT, where the circles represent dimension-6 SMEFT contributions.}
\label{fig:feynlo}
\end{figure}

\begin{figure}
    \centering
    \begin{fmffile}{nlo_box}
\begin{fmfgraph*}(100,80)
    \fmfleft{i1,i2,i3}
    \fmfright{o1,o2,o3}
    \fmftop{t1,t2}
    \fmfbottom{b1,b2}
    \fmf{dashes}{i2,v1} 
    \fmf{phantom,t=0.9}{v1,v2}
    \fmf{fermion}{v2,o2}
    \fmffreeze
    \fmf{phantom,t=1}{vtl,i3} 
    \fmf{phantom,t=0.6}{vtl,vtr} 
    \fmf{phantom,t=1}{vtr,o3} 
    \fmf{phantom,left=0.26,t=1.5}{v1,vtl,vtr} 
    \fmf{fermion,left=0.26,t=1.5}{vtr,v2} 
    \fmffreeze
    \fmf{boson,left=0.57,t=1,l.d=3}{v1,vtr} 
    \fmf{phantom,t=1}{vbl,i1} 
    \fmf{phantom,t=0.6}{vbl,vbr} 
    \fmf{phantom,t=1}{vbr,o1} 
    \fmf{phantom,right=0.26,t=1.5}{v1,vbl,vbr} 
    \fmf{boson,right=0.26,t=1.5,l.d=3}{vbr,v2}
    \fmffreeze
    \fmf{boson,right=0.57,t=1}{v1,vbr}
    \fmfblob{7}{vbr}
    \fmf{boson,t=0}{vbr,o1}
    \fmf{fermion,t=0}{o3,vtr}
    \fmfv{l.d=10,l.a=-110,l=\color{black}{$C_W$}}{vbr}
  \end{fmfgraph*}
\end{fmffile} \hspace{1cm}
\begin{fmffile}{nlo_cphi}
  \begin{fmfgraph*}(100,80)
    \fmfstraight
    \fmfright{o4,o3,o2,o1}
    \fmfleft{i2,h,i1}
    \fmftop{top}
    \fmfbottom{bot}
    \fmf{dashes,tension=1}{h,t1}
    \fmf{dashes,tension=.6}{t1,t3}
    \fmf{boson,tension=.6}{t3,t2}
    \fmf{dashes,tension=.6}{t2,t1}
    \fmf{phantom,tension=.8}{i1,t2}
    \fmf{phantom,tension=0.08}{i1,t1}
    \fmf{phantom,tension=.8}{t3,i2}
    \fmf{boson,tension=2.6}{t2,v1}
    \fmf{boson,tension=2.6}{v2,t3}
    \fmf{boson,tension=2.6}{v2,o4}
    \fmfshift{8 down}{o1}
    \fmfshift{ 4 down}{o2}
    \fmfshift{ 4 up}{o3}
    \fmfshift{8 up}{o4}
    \fmf{fermion,tension=1.3}{o1,v1,o2}
    \fmfblob{7}{t1}
    \fmfv{l.d=10,l.a=-110,l=\color{black}{$C_\phi$}}{t1}
  \end{fmfgraph*}\hspace{1cm}
\end{fmffile}
\begin{fmffile}{nlo_ceett}
  \begin{fmfgraph*}(100,80)
    \fmfstraight
    \fmfright{o4,o3,o2,o1}
    \fmfleft{i2,h,i1}
    \fmftop{top}
    \fmfbottom{bot}
    \fmf{dashes,tension=1}{h,t1}
    \fmf{fermion,tension=.6}{t1,t3,t2,t1}
    \fmf{phantom,tension=.8}{i1,t2}
    \fmf{phantom,tension=.8}{t3,i2}
    \fmfshift{8 down}{o1}
    \fmfshift{ 4 down}{o2}
    \fmfshift{ 4 up}{o3}
    \fmfshift{8 up}{o4}
    \fmf{fermion,tension=1}{o1,t2,o2}
    \fmf{phantom,tension=1}{o3,t3}
    \fmf{boson,tension=1}{o4,t3}
    \fmfblob{7}{t2}
    \fmfv{l.d=12,l.a=110,l=\color{black}{$C_{eu}[1133]$}}{t2}
  \end{fmfgraph*}
\end{fmffile}
    \caption{Representative diagrams  from SMEFT dimension-6 operators that contribute  at one-loop, but  not  at LO. The circles represent contributions proportional to the SMEFT coefficients shown in the figure.}
    \label{fig:diags_virt_only}
    \label{fig:feynnlo}
\end{figure}
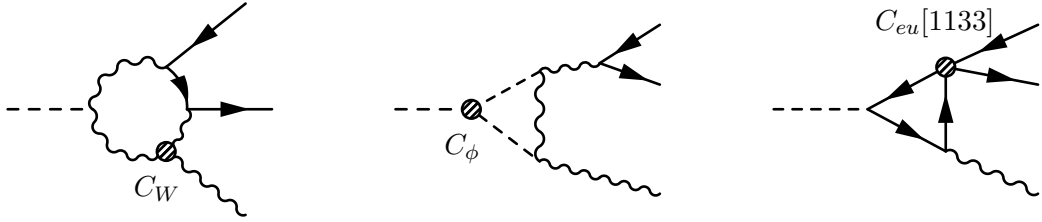

Feynman diagrams contributing to the tree level SMEFT amplitude are shown in Fig.~\ref{fig:feynlo}.  The 4-point vertex ($\ell^+\ell^- ZH$) is specific to the SMEFT, as is the $Z\gamma H$ vertex.  
Sample NLO virtual contributions are shown in Fig.~\ref{fig:feynnlo}, where we have illustrated the novel contributions to the Higgs tri-linear vertex, to the triple gauge boson vertices, and the new structure resulting from 4-fermion top quark operators. 
There are $66$ CP conserving dimension-6 SMEFT operators contribute to the NLO result
and the combinations of operators that $H\rightarrow  \ell^+\ell^- Z$ depends on are given in the appendix.

At NLO, the calculation of the virtual contributions to $H\rightarrow  \ell^+\ell^- Z$, ($\ell=e,\mu)$, is performed using the FeynRules~\cite{Alloul:2013bka} $\rightarrow $ FeynCalc~\cite{MERTIG1991345,Shtabovenko:2020gxv} $\rightarrow $ Package X~\cite{Patel:2016fam}/ Looptools \cite{HAHN1999153}/Collier~\cite{Denner:2016kdg} pipeline.    
The dimension-6 coefficients are renormalized in ${\overline{\text{MS}}}$, using the results of~\cite{Jenkins:2013zja,Jenkins:2013wua,Alonso:2013hga}, while the gauge boson masses are renormalized on-shell.
All fermions except for the top quark are consistently treated as massless in computing the virtual corrections.

The leading order and one-loop virtual results for $H(p_H)\rightarrow \ell^+(p_{e^+})+ \ell^-(p_{e^-}) +Z(p_Z)$ can be found from those for\footnote{$p_H$ is incoming, while $p_{e^+},p_{e^-}$ and $p_Z$ are outgoing.  In the scattering process, $p_1$ and $p_2$ are incoming, while $p_3$ and $p_4$ are outgoing.}
\begin{equation}
\label{eq:higgs}
 e^+(p_1)+e^-(p_2)\rightarrow Z(p_3)+H(p_4)\, ,
\end{equation}
expressed in terms of the usual Mandelstam variables,
$s=(p_1+p_2)^2$, $t=(p_1-p_3)^2$, $u=(p_3-p_1)^2$. 
Analytic results for the UV renormalized one-loop contributions to the Higgstrahlung process of eq.~\eqref{eq:higgs} can be found at~\cite{GITLAB:hllz}. To obtain the crossed result for the Higgs decay, the replacements 
\begin{equation}
 s\rightarrow m_{ee}^2=(p_{e^+} +p_{e^-})^2,\, t\rightarrow m_{13}^2=(p_{e^+}+p_Z)^2, \, u\rightarrow m_{23}^2=(p_{e^-}+p_Z)^2\, ,  
\end{equation}
must be made. 
The $H\rightarrow \mu^+\mu^- Z$ amplitudes can then be obtained by consistently swapping lepton flavor indices $1\leftrightarrow 2$ in all coefficients.
 The virtual contribution is UV finite, but contains IR poles from diagrams of purely electromagnetic origin that are canceled by real photon emission.   Our results for the decay width are consistently expanded  to ${\cal{O}}({1 / \Lambda^2})$ in the SMEFT expansion.

The $1\rightarrow 3$ amplitude is written schematically as,
 \begin{equation}
 \mathcal{A}_3=\mathcal{A}_\text{LO}+\mathcal{A}_V\, ,
 \end{equation}
 where
 we  denote the $1\rightarrow 3$ leading order matrix element as $\mathcal{A}_\text{LO}$ and the one-loop virtual contribution as $\mathcal{A}_V$. 
 The corresponding contribution to the matrix element squared is
 \begin{equation}
     |\mathcal{A}_3|^2 =|\mathcal{A}_\text{LO}|^2 +2 \text{Re}\left(\mathcal{A}_\text{LO}\mathcal{A}_{V}^*\right) \, ,
 \end{equation}
from which the width follows from the usual $1\rightarrow 3$ phase space integration.

Infrared singularities arise from the real photon emission contribution $H\rightarrow \ell^+\ell^- Z\gamma$,
and are isolated using dimensional regularization in $d=4-2\epsilon$ dimensions and  dipole subtraction techniques~\cite{Catani:1996vz,Dittmaier:2008md,Denner:2019vbn}.
We  denote the  $1\rightarrow 4$ real emission matrix element as $\mathcal{A}_4$.
All integrands are implicitly expanded  to order $\mathcal{O}(1/\Lambda^2)$.
The soft and collinear singularities are regulated by subtracting a function  that has the same IR pole structure as $\mathcal{A}_4$, then adding back the contribution after integrating analytically,
\begin{eqnarray}
    \int d\Gamma_4 &= 
    & 
    \frac{1}{2m_H} \int d\Phi_4 \left(|\mathcal{A}_4|^2 - |\mathcal{A}_\text{sub}|^2 
    \right) + \frac{1}{2 m_H} \int d\Phi_3 
    \left[\int dp_\gamma \otimes |\mathcal{A}_\text{sub}|^2
    \right] \, .
\end{eqnarray}
Here the integrals over $d\Phi_3$ and $d\Phi_4$ indicate integration over the three-body and four-body phase space, respectively, 
and
\begin{eqnarray}
| \mathcal{A}_\text{sub}|^2&=& 2 e^2{p_{\ell^+}\cdot p_{\ell^-}\over ( p_{\ell^+}\cdot p_\gamma)(p_{\ell^-}\cdot p_\gamma)}\, .
\end{eqnarray}

In ${\overline{\text{MS}}}$, the integrated subtraction term is~\cite{Denner:2019vbn}
\begin{eqnarray}
&& \frac{1}{2m_H}\int d\Phi_3 \left[\int dp_\gamma \otimes |\mathcal{A}_\text{sub}|^2 \right] =  
\nonumber \\
&& 
 \frac{\alpha}{4\pi m_H} 
 \bigg[\int d\widetilde{\Phi}_3 |\mathcal{A}_\text{LO}|^2 \int_0^1 dz \left\{G^{(\text{sub})}({\widetilde{m}}^2_{\ell\ell})\delta{(1-z)} 
 +\left[\bar{\mathcal{G}}
 ({\widetilde{m}}_{\ell\ell}^2,z)\right]_+\right\} 
 \nonumber \\ &&
    \times\Theta_\text{cut}(p_{\ell^-} = z\widetilde{p}_{\ell^-},p_{\ell^+} = \widetilde{p}_{\ell^+},p_\gamma = (1-z)\widetilde{p}_{\ell^-})\bigg]+(p_{\ell^-} \leftrightarrow p_{\ell^+}) \, ,
\end{eqnarray}
where ${\widetilde{m}}_{\ell\ell}^2 \equiv (\widetilde{p}_{\ell^-}+\widetilde{p}_{\ell^+})^2$ is independent of the $z$ integral,\footnote{The tilde over $d\widetilde{\Phi}_3$ is to emphasize that this integration is over the lepton momenta $\widetilde{p}_{\ell^+}$ and $\widetilde{p}_{\ell^-}$.} the $[]_+$ indicates the usual plus distribution, $\int dz[\bar{\mathcal{G}}(z)]_+ \Theta(z) = \int dz \bar{\mathcal{G}}(z) [\Theta(z)-\Theta(1)]$, and $\alpha$ is defined in  using the SMEFT relations of eq.~\eqref{eq:parshift}.  
The function $\Theta_\text{cut}$ indicates the definitions of the momenta $p_{\ell^-}$, $p_{\ell^+}$, and $p_\gamma$ that are subject to phase space cuts, which are \textit{not} the same as the original four-body phase space momenta~\cite{Dittmaier:2008md,Denner:2019vbn}. 
Note that for fully inclusive observables with no phase space cuts such as the total width, $\Theta(z)=\Theta(1)$ and so $\bar{\mathcal{G}}$ does not contribute.
The functions $G$ and $\bar{\mathcal{G}}$ are given by~\cite{Denner:2019vbn,Dittmaier:2008md}
\begin{align}
\begin{split}
    G({\widetilde{m}}_{\ell\ell}^2) =& \Gamma(1+\epsilon)\left(\frac{4\pi\mu^2}{{\widetilde{m}}_{\ell\ell}^2}\right)^\epsilon \left(\frac{1}{\epsilon^2}+\frac{3}{2\epsilon}\right)+\frac{7}{2}-\frac{\pi^2}{3}\\
    \bar{\mathcal{G}}({\widetilde{m}}_{\ell\ell}^2,z) =& 
    \hat{P}_{ff}
    \left[\frac{-(4\pi)^\epsilon}{\epsilon \Gamma(1-\epsilon)} + \ln\left(\frac{{\widetilde{m}}_{\ell\ell}^2}{\mu^2}\right) + \ln z +2 \ln(1-z)\right] + (1+z)\ln(1-z)+1-z
\end{split}
\end{align}
where $\hat{P}_{ff} =(1+z^2)/(1-z)$,  we expand in $\epsilon$ and drop all terms of $\mathcal{O}(\epsilon)$, and $\mu$ is an arbitrary renormalization scale.
The IR singularities in the virtual contributions cancel with the singularities in $G$.
However, additional collinear singularities proportional to $\epsilon^{-1} \hat{P}_{ff}$ appear in non-inclusive observables from real emission, which are encoded in $\bar{\mathcal{G}}$.
This divergence can be reabsorbed by expressing $\bar{\mathcal{G}}$ using the lepton mass as a regulator by applying the techniques of \cite{Denner:2019vbn},
\begin{equation}
    \bar{\mathcal{G}}_\text{MR}({\widetilde{m}}_{\ell\ell}^2,z) = \hat{P}_{ff}\left[\ln{\left(\frac{{\widetilde{m}}_{\ell\ell}^2}{m_\ell^2}\right)}+\ln{z}-1\right]+(1+z)\ln{\left(1-z\right)}+1-z \, ,
\end{equation}
where $m_\ell$ is the lepton mass, thus replacing the divergence in $\epsilon\to 0$ with a logarithmic divergence in the mass of the lepton, which plays the role of a physical cutoff.

\section{Results}
\label{sec:results}
The experimental values of the input parameters are,
\begin{equation}
\begin{aligned}[c]
m_W^\text{exp}&=80.379 \ \text{GeV}\, ,\\
m_Z^\text{exp}&=91.1876 \ \text{GeV}\, ,\\
m_H&= 125.1 \ \text{GeV}\, ,\\
m_\mu &= 105.7 \ \text{MeV} \, ,
\end{aligned}
\qquad\qquad \qquad
\begin{aligned}[c]
G_\mu&=1.16638 \times 10^{-5} \ \text{GeV}^{-2}\, ,\\
m_t&= 172.76 \ \text{GeV}\, ,\\
m_e&= 0.511 \ \text{MeV} \, ,
\end{aligned}
\end{equation}
where the lepton masses only enter in the  logarithmic corrections coming from real photon emission. 
The masses $m_W$ and $m_Z$ that we use to derive  our results are 
\begin{align}
\label{eq:mods}
\begin{split}
    m_Z &= \frac{m_Z^\text{exp}}{\sqrt{1+\left(\Gamma_Z^\text{exp}/m_Z^\text{exp}\right)^2}} = 91.1535 \ \text{GeV} \, ,\\
    m_W &= \frac{m_W^\text{exp}}{\sqrt{1+\left(\Gamma_W^\text{exp}/m_W^\text{exp}\right)^2}} = 80.352 \ \text{GeV} \, ,
\end{split}
\end{align}
where the modifications of eq.~\eqref{eq:mods} approximate finite width effects~\cite{Bardin:1988xt}.

\subsection{Standard Model Results}

\begin{figure}
    \centering  \includegraphics[width=.6\linewidth]{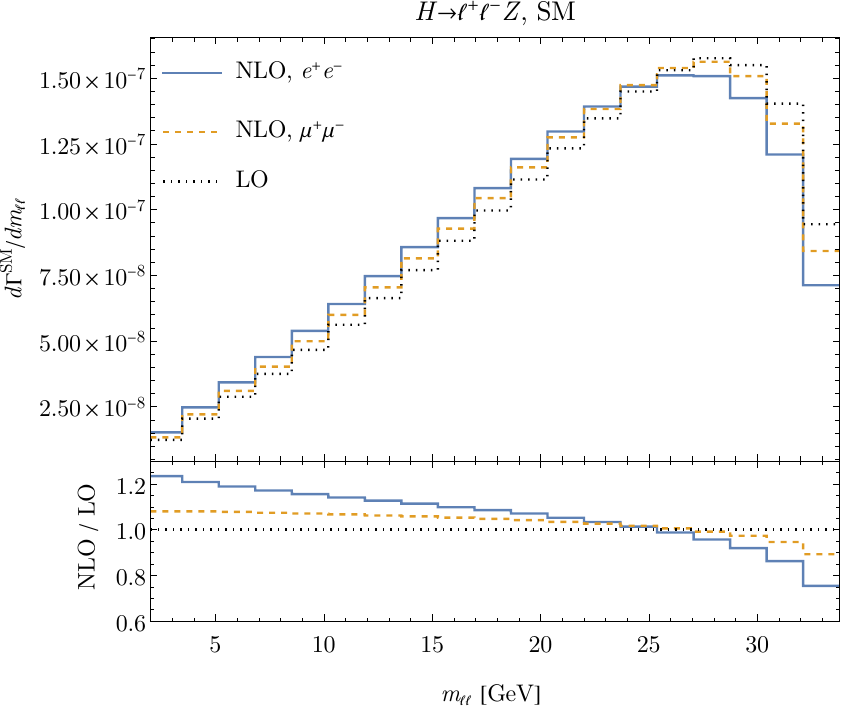}
    \caption{The impact of the NLO corrections on the differential decay width $d\Gamma^\text{SM}/dm_{\ell\ell}$ in the Standard Model. }
    \label{fig:sm}
\end{figure}

The distribution of the produced lepton pair in the SM is shown in Fig. \ref{fig:sm} and the NLO effects are of ${\cal{O}}(20\%)$ at both high and low $m_{\ell\ell}$.  We observe that the distributions for $e^+e^- Z$ and $\mu^+\mu^- Z$ are slightly different due to $\log\left({m_{\ell\ell}^2 / m_e^2}\right)$ and 
$\log\left({m_{\ell\ell}^2 / m^2_\mu}\right)$
 effects in the real photon emission contributions. With our inputs, the integrated SM NLO rate is $\Gamma^\text{SM}_\text{NLO}(H\rightarrow \ell^+ \ell^- Z) = 2.997\times 10^{-3}$ MeV, where the $\log({m_{\ell\ell}^2 / m_\ell^2})$ contributions cancel in the total rate.

\subsection{NLO SMEFT Rates for $H\rightarrow  \ell^+ \ell^- Z$}

The width for $H\rightarrow \ell^+ \ell^- Z$ at tree-level in the SMEFT is well known~\cite{Buchalla:2013mpa,Isidori:2013cla},
\begin{align}
\begin{split}
    \frac{\Gamma^\text{SMEFT}_\text{LO}(H\rightarrow \ell^+_i\ell^-_iZ)}{\Gamma^\text{SM}_\text{NLO}(H\rightarrow \ell^+_i\ell^-_iZ)} =& \frac{1 \ \text{TeV}^2}{\Lambda^2} \bigg[
-0.0851 C_{\phi B}
+0.0199 C_{\phi D}
+0.119 C_{ \phi \square }
+0.0268 C_{\phi W} \\
&-0.0510 C_{\phi WB}
-0.00691 C_{\phi e}[ii]
+0.00859 C_{\phi l}^{(1)}[ii]\\
&-0.111 C_{\phi l}^{(3)}[ii]
-0.119 C_{\phi l}^{(3)}[jj]
+0.119 C_{ll}[1221]
    \bigg] \, ,
\end{split}
\end{align}
where  we used the flavor index $i$ to refer to the lepton produced in the decay ($\ell_i=(e,\mu)$) and the flavor index $j$ to refer to the lepton not produced in the decay ($\ell_j = (\mu,e)$).

\begin{figure}
    \centering 
\includegraphics[width=0.49\linewidth]{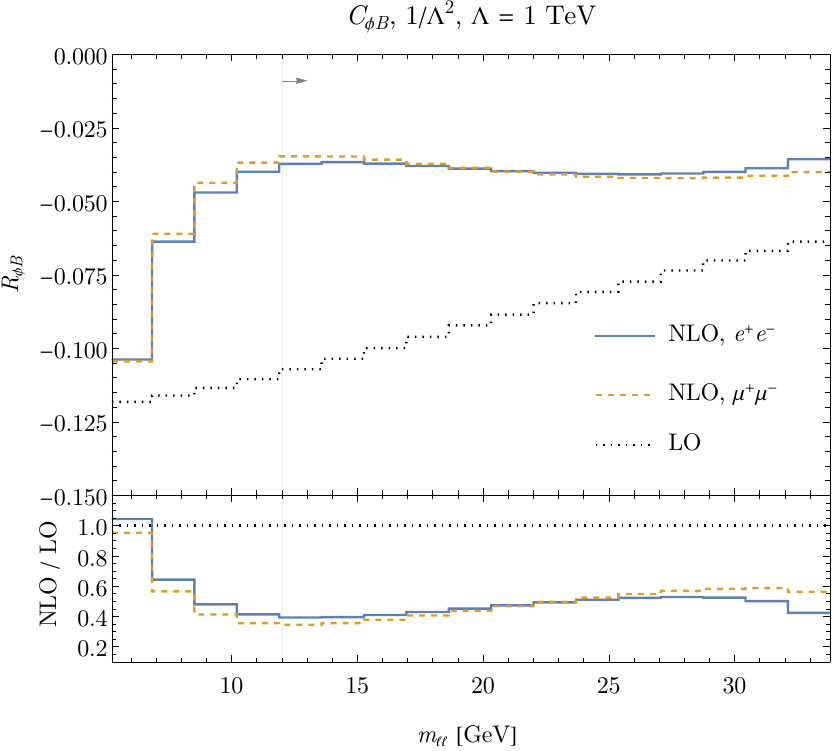}
\includegraphics[width=0.49\linewidth]{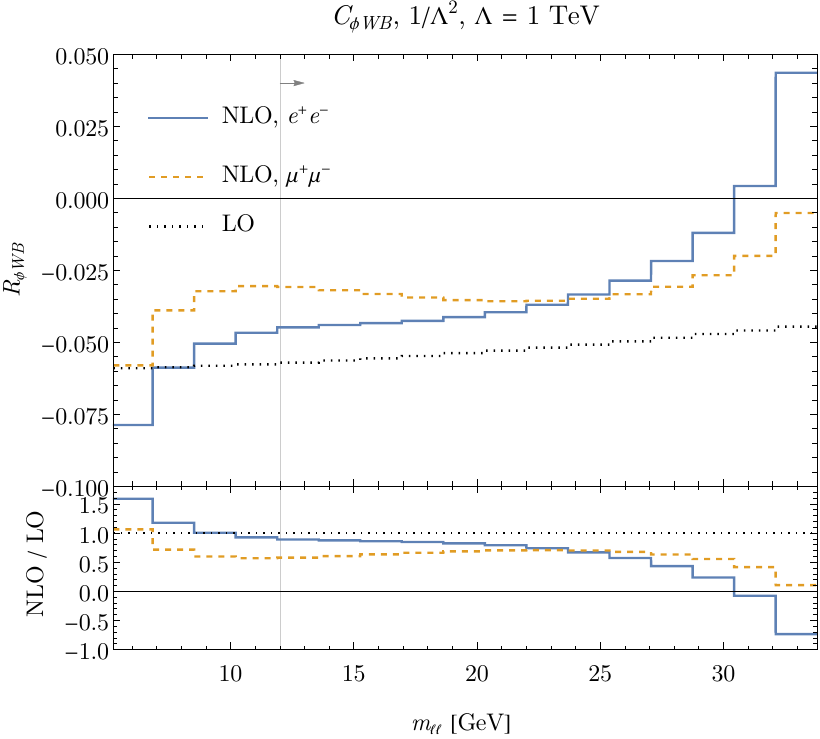}
    \caption{The contribution of the coefficients $C_{\phi B}$ (left) and $C_{\phi WB}$ (right) to the differential decay width at LO and NLO, normalized to the SM contribution with the definitions of eqs.~\eqref{eq:rdef} and~\eqref{eq:ratdef}.  The vertical line at $12~\text {GeV}$ shows the typical experimental cut.}
    \label{fig:CphiWBrats}
\end{figure}

The SMEFT contributions change the shape of the $m_{\ell\ell}$ distributions, as illustrated in Fig.~\ref{fig:CphiWBrats}. 
We write the $m_{\ell\ell}$ SMEFT distributions as,
\begin{eqnarray}
{d \Gamma_\text{NLO}\over dm_{\ell\ell}}&=&{d \Gamma_\text{NLO}^\text{SM}\over dm_{\ell\ell}}+\sum_{i}
{C_i\over\Lambda^2}{d\Gamma_{i,\text{NLO}}^\text{SMEFT}\over dm_{\ell\ell}}\, ,
\end{eqnarray}
where $\Gamma_\text{NLO}$ contains both the LO and the NLO contributions.
The upper portion of Fig. \ref{fig:CphiWBrats} is
\begin{eqnarray}
R_{i,\text{NLO}}&=& (1~\text{TeV})^2 {d\over dC_i}\biggl[{d\Gamma_\text{NLO}\over dm_{\ell\ell}}
\biggr]/\biggl[
 {d \Gamma^{SM}_\text{NLO}\over dm_{\ell\ell}}\biggr]\nonumber \\
R_{i,\text{LO}}&=&
(1~\text{TeV})^2 {d\over dC_i}\biggl[{d\Gamma_\text{LO}\over dm_{\ell\ell}}
\biggr]/\biggl[
 {d \Gamma^\text{SM}_\text{LO}\over dm_{\ell\ell}}\biggr]
\label{eq:rdef}
\end{eqnarray}
The lower portion of Fig. \ref{fig:CphiWBrats} shows the relative effect of the NLO contributions for specific operators, 
\begin{eqnarray}
{\text{NLO}\over \text{LO}}&=&
{d\over dC_i}\biggl[{d\Gamma_\text{NLO}\over dm_{\ell\ell}}
\biggr]/{d\over dC_i}\biggl[{d\Gamma_\text{LO}\over dm_{\ell\ell}}
\biggr]
\label{eq:ratdef}
\end{eqnarray}

For $\mathcal{O}_{\phi B}$, the NLO corrections suppress the SMEFT contribution at larger values of $m_{\ell\ell}$ while
the corrections from $\mathcal{O}_{\phi W B}$ enhance the rate at small $m_{\ell\ell}$, but suppress it for $m_{\ell\ell} > 12~\text{GeV}$. For both of these operators, we see that the NLO corrections significantly change the shape of the distribution, due to the presence of new kinematic structures.

\begin{figure}
    \centering
    \includegraphics[width=0.46\linewidth]{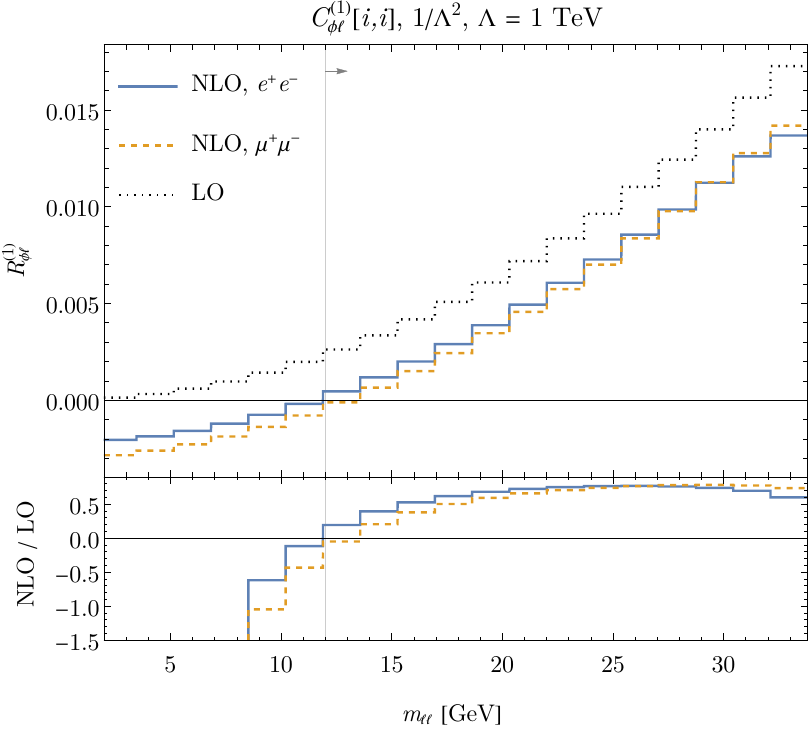}
    \includegraphics[width=0.49\linewidth]{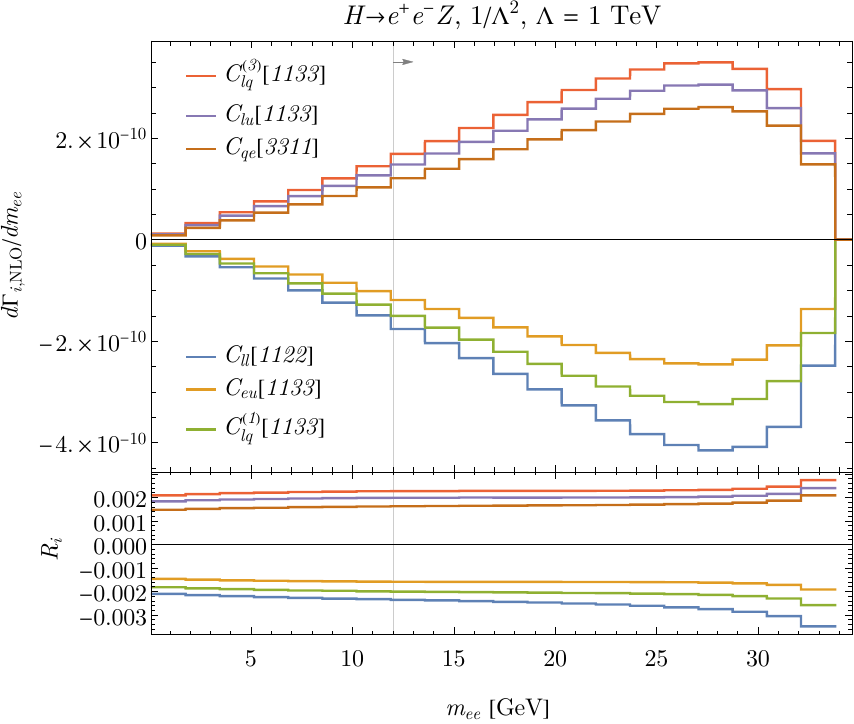}
    \caption{LHS.  Contributions from 2-fermion operators that contribute at LO  with the definitions of eqs.~\eqref{eq:rdef} and~\eqref{eq:ratdef}.  The vertical line at $12~\text {GeV}$ shows the typical experimental cut.  RHS:  Contributions from 4-fermion operators involving top quark loops that first appear at NLO.}
    \label{fig:Cfl1}
\end{figure}

Fig.~\ref{fig:Cfl1}
shows the $m_{\ell\ell}$ distributions resulting from representative 2-fermion and 4-fermion operators.  The 4-fermion operators shown involve top quark loops that first occur at NLO and are enhanced at large $m_{\ell\ell}$, while the 2-fermion operators suppress the rate relative to the LO.

\begin{figure}
    \centering 
    \includegraphics[width=0.65\linewidth]{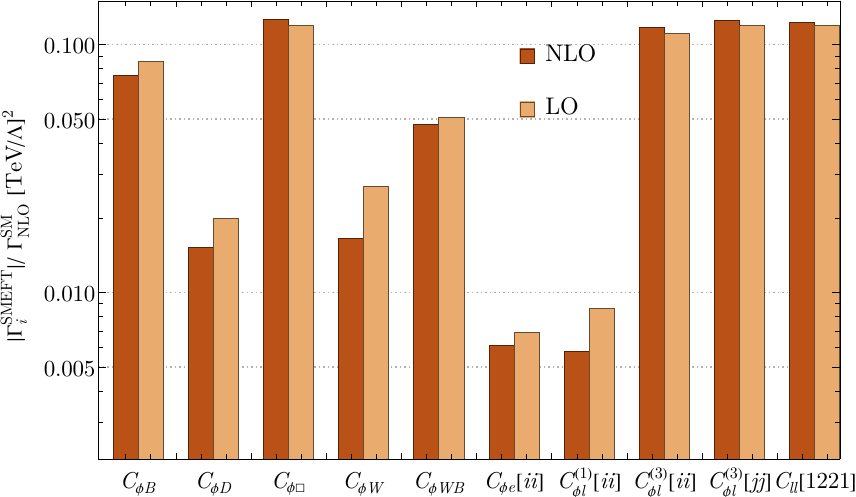}
    \caption{Impact of NLO corrections on the contributions of operators that occur at LO in the decay $H\rightarrow \ell^+\ell^- Z$. The $y$-axis is defined in eq.~\eqref{eq:difdef}  and the numerical values are given in Table \ref{tab:kfactors}. The flavor index $i$ corresponds to the produced lepton $(e,\mu)$, and $j$ is the lepton that is not produced, $(\mu,e$).}
    \label{fig:bar_chart_incl}
\end{figure}

Our results for the  total width for $H\rightarrow  \ell^+\ell^- Z$ are presented as a 
series of tables.  The NLO prediction for the decay width $H\rightarrow  \ell^+\ell^- Z$ is  parameterized as,
\begin{eqnarray}
    \Gamma_\text{NLO}(H\rightarrow  \ell_i^+\ell_i^- Z) &=&\Gamma^\text{SM}_\text{NLO}(H\rightarrow   \ell_i^+\ell_i^- Z)
    +\delta \Gamma^\text{SMEFT}(H\rightarrow \ell_i^+\ell_i^- Z)\nonumber \\
    &=& 2.997\times 10^{-3}~\textrm{MeV}+\sum_k{\beta_k^{(l)}C_k\over\Lambda^2}\, ,
    \label{eq:smeftdef}
\end{eqnarray}
and $\beta_k^{(0)}, \beta_k^{(1)}$ are the tree level and one-loop  plus real  SMEFT contributions. 
 Table \ref{tab:kfactors}
contains the effects of operators that contribute at LO and the results are summarized in Fig. \ref{fig:bar_chart_incl}.  The y-axis is of Fig. \ref{fig:bar_chart_incl} is,
\begin{eqnarray}
\label{eq:difdef}
{| \Gamma_{i, \text{NLO}}^\text{SMEFT}|\over \Gamma_\text{NLO}^\text{SM}}&=&
\left( {1~\text{TeV}^2\over \Lambda^2}\right){|\beta_i^{(0)}+\beta_i^{(1)}|\over \Gamma_\text{NLO}^\text{SM}}\nonumber \\
{| \Gamma_{i, \text{LO}}^\text{SMEFT}|\over \Gamma_\text{NLO}^\text{SM}}&=&
\left( {1~\text{TeV}^2\over \Lambda^2}\right){|\beta_i^{(0)}|\over \Gamma_\text{NLO}^\text{SM}}\, .
\end{eqnarray}
   Tables \ref{tab:virtonly_2f} and \ref{tab:virtonly_4f} contain the numerical contributions to the total width from operators that first arise at NLO.  

\begin{figure}
    \centering
    \includegraphics[width=0.65\linewidth]{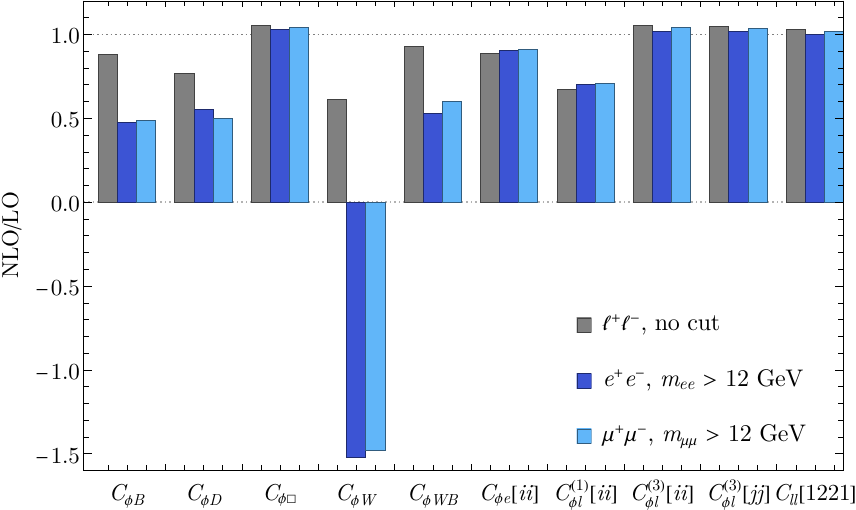}
    \caption{Impact of NLO corrections on the contributions of operators that occur at LO in the decay $H\rightarrow \ell^+\ell^- Z$ and demonstrating the effect of the cut on $m_{\ell\ell}$.
    With a small abuse of notation, in this figure NLO/LO is defined as an inclusive analogue of eq.~\eqref{eq:ratdef}, i.e. $(\beta_k^{\left(0\right)}+\beta_k^{\left(1\right)})/\beta_{k}^{\left(0\right)}$ and $(\widetilde{\beta}_k^{\left(0\right)}+\widetilde{\beta}_k^{\left(1\right)})/\widetilde{\beta}_{k}^{\left(0\right)}$ for the gray and blue columns, respectively, where $\beta_k^{(l)}$ is defined in eq.~\eqref{eq:smeftdef}.}
    \label{fig:bar_chart_mll_cut}
\end{figure}

\subsection{NLO SMEFT Rates for $H\rightarrow  \ell^+ \ell^- Z$ with $m_{\ell\ell}$ cuts}

Since CMS and ATLAS generally impose a cut of $m_{\ell\ell}>12$ GeV in their measurements of the $H\rightarrow$ 4 lepton width~\cite{CMS:2021ugl,ATLAS:2020rej}, we also provide results with this cut. 
We denote this width as $\widetilde{\Gamma} \equiv \int_{m_{\ell\ell}>12~\textrm{GeV}} d\Gamma$ in the following. 
After imposing this cut, the logarithms of the lepton mass that canceled in fully inclusive measurements no longer cancel, and so the $e^+e^-$ and $\mu^+\mu^-$ channels differ at NLO due to the real emission contributions.
In the SM, with our inputs, we find 
\begin{align}
\begin{split}
    \widetilde{\Gamma}^\text{SM}_\text{NLO}(H\rightarrow e^+e^- Z) = 2.582 \times 10^{-3}~\text{MeV}\, ,\\
    \widetilde{\Gamma}^\text{SM}_\text{NLO}(H\rightarrow \mu^+\mu^- Z) = 2.616 \times 10^{-3}~\text{MeV} \, .
\end{split}
\end{align}

For those operators first appearing at one-loop, there are no real emission contributions and so  the widths for the $e^+e^-$ and $\mu^+\mu^-$ channels are the same at this order.
We write these contributions in Tables~\ref{tab:virtonly_2f_qcut} and~\ref{tab:virtonly_4f_qcut}. The effect of this cut is to significantly enhance the relative importance of the NLO contributions for many of the operators, in some cases by as much as a factor of 2.  This can be seen  by comparing the  far right column of  Table \ref{tab:kfactors} with Tables  \ref{tab:kfactors_q12_ee} and \ref{tab:kfactors_q12_mumu}.
The importance of the experimental cut on $m_{\ell\ell}$ is seen clearly in Fig. 
\ref{fig:bar_chart_mll_cut}.

\subsection{Narrow Width Approximation to $H\rightarrow \ell_i^+\ell_i^- Z, \ Z\rightarrow \ell_j^+\ell_j^-$}\label{sec:NW}

\begin{figure}
    \centering
    \includegraphics[width=0.49\linewidth]{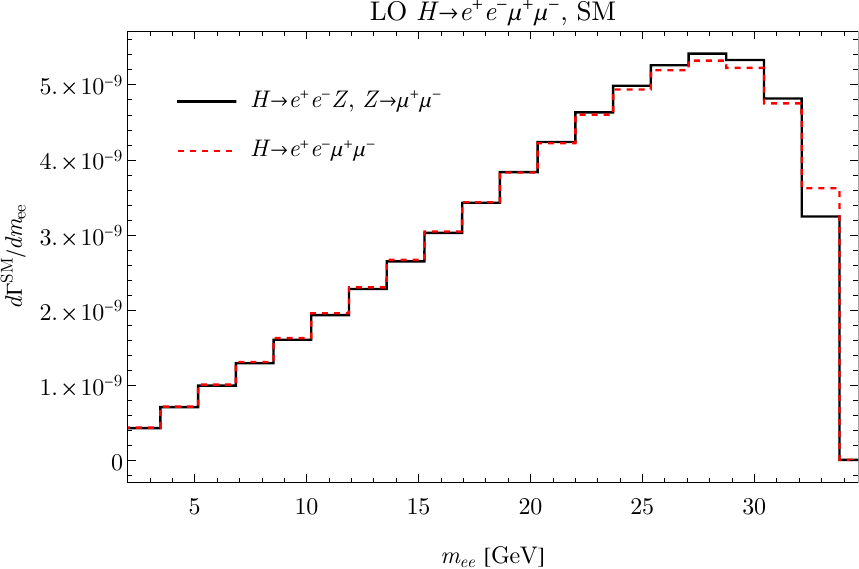}
    \includegraphics[width=0.49\linewidth]{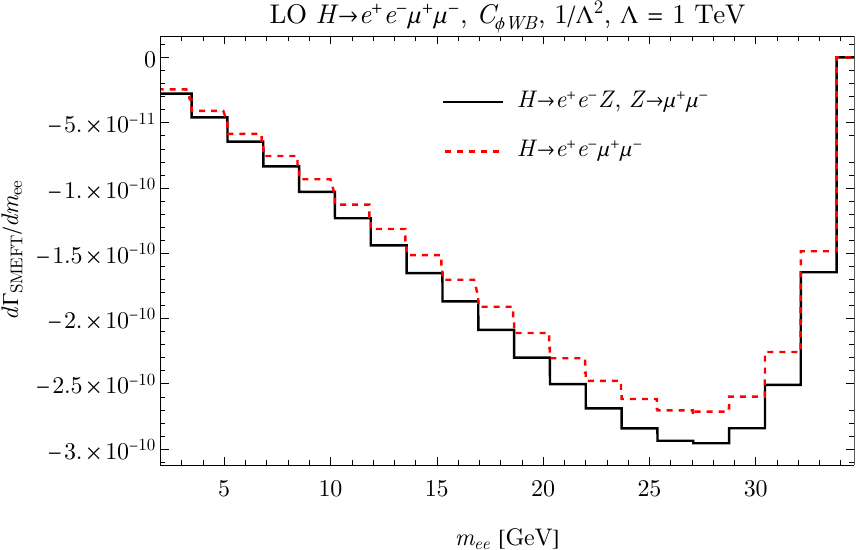}
    \caption{LO comparison of the full four-body $H\rightarrow e^+e^-\mu^+\mu^-$ and the narrow width approximated $H\rightarrow e^+ e^- Z, \ Z\rightarrow \mu^+\mu^-$ differential rates for the SM and for the contribution proportional to $C_{\phi WB}$.}
    \label{fig:Full_1to4_Comparison}
\end{figure}
We use the narrow width approximation to compute the decay of $H\rightarrow \ell_i^+\ell_i^- Z, \ Z\rightarrow \ell_j^+\ell_j^-$  at NLO electroweak order in the SMEFT,
\begin{equation}
\Gamma(H\rightarrow \ell_i^+\ell_i^-\ell_j^+\ell_j^-)=(2-\delta_{ij})\Gamma(H\rightarrow \ell_i^+\ell_i^- Z)\text{BR}(Z\rightarrow \ell_j^+\ell_j^-)\, .
\end{equation}
The NLO branching ratio (BR) of $Z$ to leptons is parameterized as
\begin{eqnarray}
    \text{BR}(Z\rightarrow \ell_i^+\ell_i^-) &=&\text{BR}^\text{SM}(Z\rightarrow \ell_i^+\ell_i^-)
    +\delta \text{BR}^\text{SMEFT}(Z\rightarrow \ell_i^+\ell_i^-)\, ,\nonumber \\
    &=& 0.033670+\sum_k{\alpha_k^{(l)}C_k\over\Lambda^2}
    \label{eq:brzee}
\end{eqnarray}
where have  inserted the most accurate theoretical prediction~\cite{Freitas:2014hra} for the SM contribution in eq.~\eqref{eq:brzee}.  
Numerical values for the $\alpha_k^{(l)}$ at both LO and NLO can be found  using the results of~\cite{Dawson:2019clf,Dawson:2022bxd,Bellafronte:2023amz}.

At LO, it is straightforward to assess the accuracy of the narrow width approximation, both in the SM and in the SMEFT.  In Fig. \ref{fig:Full_1to4_Comparison}, we show the invariant mass distribution of the $e^+e^-$ pair for $H\rightarrow  e^+e^- \mu^+\mu^-$ for the complete four-body decay\footnote{We emphasize that this includes {\bf{all}} leading order contributions, including $H\rightarrow Z\gamma\rightarrow \ell^+\ell^-\ell^+\ell^-$.} and in the narrow width approximation for the SM (LHS) and the contribution from a representative SMEFT coefficient (RHS)\footnote{The narrow width result shown in Fig.~\ref{fig:Full_1to4_Comparison} uses LO predictions everywhere in order to be self-consistent.}. 
 In general, the narrow width approximation for Higgs decays in the SMEFT  could fail due to contributions from interactions in the SMEFT that are not present in the SM~\cite{Brivio:2019myy}.  However, at LO the narrow width approximation is extremely accurate for the operators that contribute to the $H\rightarrow$ 4 lepton process.  This motivates our use of the narrow width approximation in our NLO SMEFT calculation.

\begin{figure}
    \centering
    \hskip-.1in\includegraphics[width=0.479\linewidth]{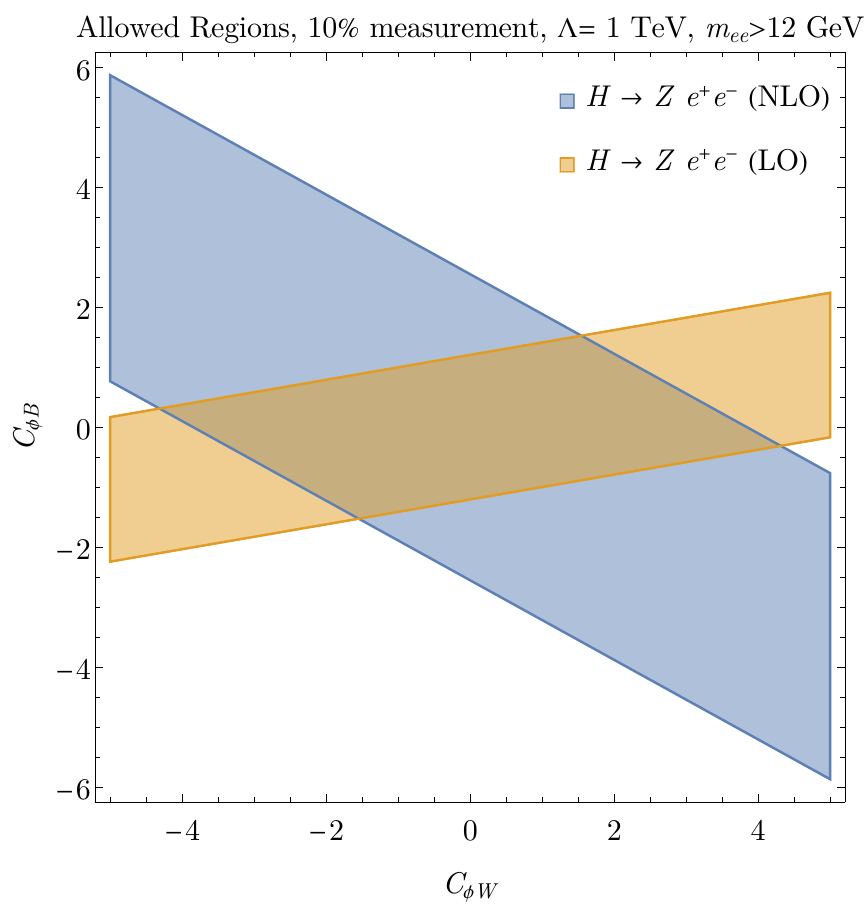}
    \hskip.1in\includegraphics[width=0.49\linewidth]{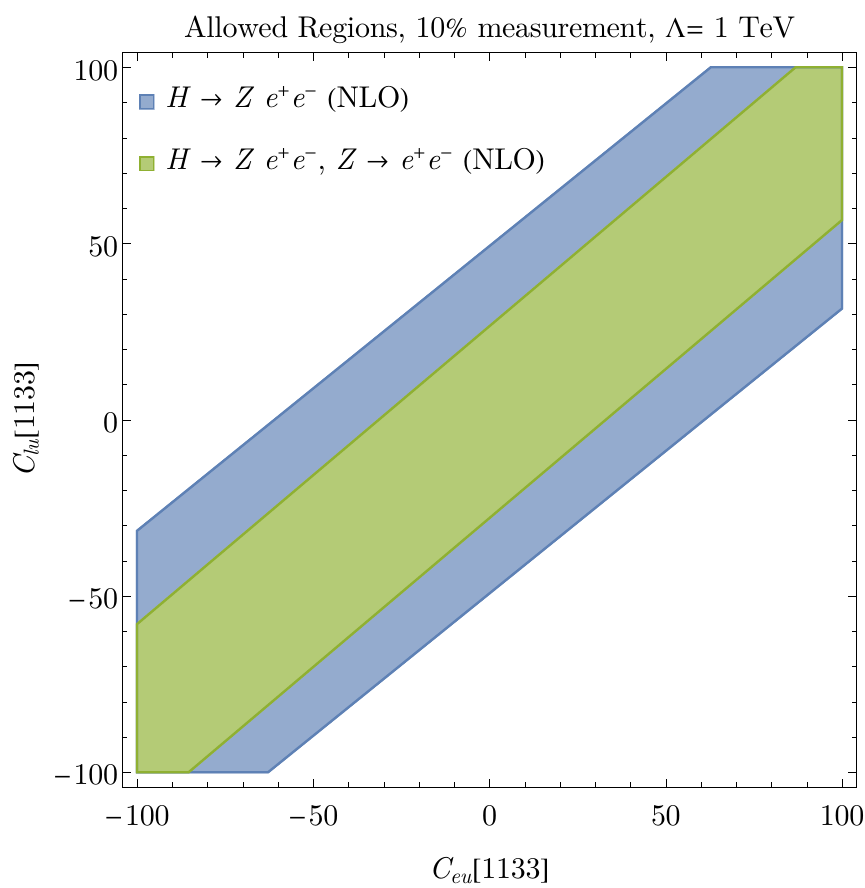}
    \caption{Regions where $\widetilde{\Gamma}(H\rightarrow  e^+e^- Z)$ is within  $10\%$ \cite{CMS:2021ugl}  of the NLO SM prediction, applying the realistic experimental cut of $m_{ee} > 12$ GeV.
 The blue and orange curves show the NLO and LO $\Gamma(H\rightarrow  e^+e^- Z)$ results, respectively. 
 The green curve shows the NLO result for $H\rightarrow  e^+e^- Z, Z\rightarrow e^+e^-$ in the narrow width approximation.}
    \label{fig:rats}
\end{figure}

In Fig.~\ref{fig:rats}, we show the region where the prediction for $\widetilde{\Gamma}(H\rightarrow e^+e^- Z)$ in a 2-parameter SMEFT fit  is  within $10\%$\cite{CMS:2021ugl} of the NLO SM prediction, including the cut $m_{\ell\ell} > 12$ GeV.
On the LHS, we see how the NLO corrections can significantly change the correlation between operators.   
On the RHS of Fig. \ref{fig:rats}, we show the correlation between the effects of 2 operators  ($\mathcal{O}\sim {\overline{t}}t {\overline {e}}e$) that first arise at NLO and here we see the impact of the correlation between operator contributions and the significant effect of including the NLO results both in $H\rightarrow  e^+e^- Z$ and in $Z\rightarrow e^+e^-$ to obtain a consistent NLO prediction in the narrow width approximation.

\begin{figure}
    \centering
    \includegraphics[width=0.49\linewidth]{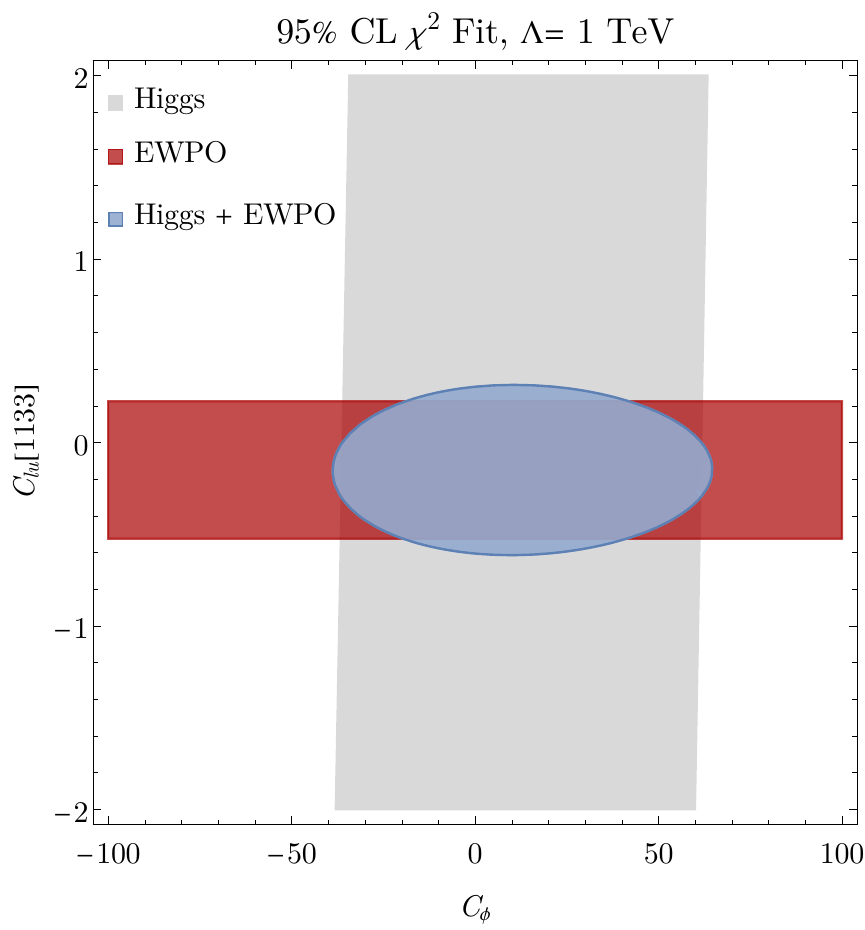}
 \includegraphics[width=0.49\linewidth]{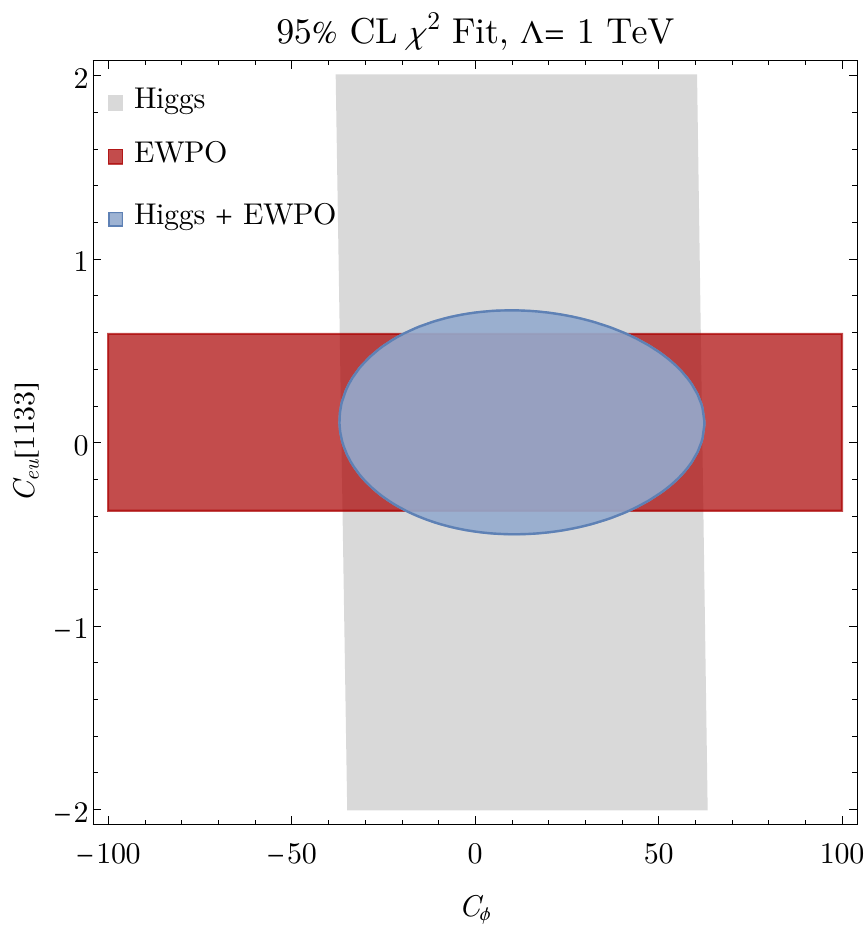}
    \caption{$\chi^2$ fit to LHC data for $H\rightarrow \ell^+\ell^- Z, Z\rightarrow \ell^+\ell^-$    and to EWPOs using NLO predictions in the narrow width approximation.} 
    \label{fig:chi2.pdf}
\end{figure}

Fig.~\ref{fig:chi2.pdf} employs Higgs data from all production channels relevant for the decay $H\rightarrow $  4 leptons, adapting the fit of~\cite{Dawson:2022cmu} for Higgs decays and the fit of~\cite{Dawson:2019clf} to include the EWPOs. The resulting predictions are parameterized as,
\begin{equation}
\mu\equiv\biggl[
{\sigma_{ij\rightarrow H}\over (\sigma_{ij\rightarrow H})\mid_\text{SM}}\biggr]
\biggl[
{\text{BR}(H\rightarrow  \ell^+\ell^- Z)\over \text{BR}(H\rightarrow \ell^+\ell^- Z) \mid_\text{SM}}\biggr]
\biggl[
{\text{BR}(Z \rightarrow \ell^+\ell^-)\over \text{BR}(Z\rightarrow  \ell^+\ell^-) \mid_\text{SM}}\biggr]\, ,
\end{equation}
where each expression in the square brackets is linearized in the dimension-6 SMEFT coefficients.  The term with $\sigma_{ij\rightarrow H}$ represents the various Higgs production channels, and we include only the $C_\phi$ contribution in this piece.
Fig.  \ref{fig:chi2.pdf}~ shows the $95\%$ CL limits on $C_\phi$ and on two 4- fermion operators involving the top quark (which are chosen to be operators that do not contribute to Higgs production) using the narrow width approximation to obtain consistent NLO fits.  The figure demonstrates the interplay of Higgs and electroweak data.

\section{Conclusions}

We computed the NLO electroweak corrections to the Higgs decays $H\rightarrow  \ell^+\ell^- Z$ and $H\rightarrow $ 4 leptons in the SMEFT. We included the contributions coming from all the dimension-6 operators, without any assumptions on their flavor structures, but dropping contributions proportional to the off-diagonal elements of the CKM matrix.  
The $H\rightarrow $ 4 lepton decay rate was calculated using a narrow width approximation, by combining the $H\rightarrow  \ell^+\ell^- Z$ rate, calculated here, and the $Z\rightarrow \ell^+\ell^-$ rate, known in the literature. For both processes, the rates are known at full NLO in the SMEFT up to dimension-6. In section~\ref{sec:NW} we show that at LO the narrow width approximation is very accurate in reproducing the contribution of the operators that affect $H\rightarrow $ 4 leptons.

The effects of the NLO SMEFT contributions can be significant and affect both the total rate and the shape of the $m_{\ell\ell}$ distributions, along with introducing a dependence on operators that do not contribute at LO. Mirroring the experimental collaborations, we notice that, by considering a lower cut on the final lepton invariant mass $m_{\ell\ell}>12$ GeV, the importance of NLO contributions to $H\rightarrow  \ell^+\ell^- Z$ is enhanced for many operators. Our numerical results demonstrate the large correlations between the effects of different operators and show that single operator fits can be significantly misleading.

These results are important for the study of Higgs physics at the LHC and future colliders, as they provide precise information on the type of new physics that is accessible in these searches. Furthermore, the calculation presented in this paper is  a fundamental component of an eventual SMEFT global fit that is accurate to NLO.  

Results for the numerical contributions to the total width for $H\rightarrow  \ell^+\ell^- Z$ are given in the appendix, while
analytic results for the virtual NLO contributions used in this work can be found at 
\cite{GITLAB:hllz}.

\section*{Acknowledgments}

We thank K. Asteriadis and R. Szafron for valuable discussions and the invaluable collaboration on \cite{Asteriadis:2024qim,Asteriadis:2024xts}.
S. D. is supported by the U.S. Department of Energy under Grant Contract~DE-SC0012704. M.F. is supported by the U.S. Department of Energy, Office of Science, Office of Workforce Development for Teachers and Scientists, Office of Science Graduate Student Research (SCGSR) program. The SCGSR program is administered by the Oak Ridge Institute for Science and Education (ORISE) for the DOE. ORISE is managed by ORAU under contract number DE-SC0014664. P.P.G. is supported by the Ramón y Cajal grant~RYC2022-038517-I funded by MCIN/AEI/10.13039/501100011033 and by FSE+, and by the Spanish Research Agency (Agencia Estatal de Investigación) through the grant IFT Centro de Excelencia Severo Ochoa~No~CEX2020-001007-S. 
\bibliographystyle{apsrev4-1}
\bibliography{hzee.bib}
\newpage
\appendix
\section{Numerical Results}

In this appendix we report the numerical results for the  contributions to the total width for $H\rightarrow  e^+ e^- Z$.  The width for $H\rightarrow  \mu^+ \mu^- Z$ is found by making the change, $1\leftrightarrow 2$, in the lepton flavor indices. At LO, 
there are 10
operators that contribute to the decay, 
\begin{align}
C_{\phi D}\, ,~C_{\phi \square}\, ,~
C_{\phi WB}\, ,~C_{\phi W}\, ,~
C_{\phi B}\, ,~C_{\phi e}[11]\,  ,C_{\phi l}^{(1)}[11]\, ,C_{\phi l}^{(3)}[11]\, ,C_{\phi l}^{(3)}[22]\, , C_{ll}[1221].
\end{align}

At NLO, an additional 20 2-fermion operators contribute to $H\rightarrow  e^+ e^- Z$,
\begin{eqnarray}
&&C_{\phi d}[ii],~ C_{\phi u}[ii],~ C_{\phi q}^{(1)}[ii],~ C_{\phi q}^{(1)}[ii], \quad i=1,2,3\nonumber \\
&&C_{\phi e}[jj],~C_{\phi l} ^{(1)}[jj], \quad j=2,3 \nonumber \\
&&
C_{\phi l}^{(3)}[33],~ C_{uW}[33],~C_{uB}[33],~C_{u\phi}[33]\, .
\end{eqnarray}
At NLO, there are 34  new 4-fermion operators that contribute to $H\rightarrow  e^+ e^- Z$, where we note that $C_{ee}$ and $C_{ll}$ obey the symmetries $C_{X}[kkii]=C_{X}[iikk]$, $C_{X}[ikki]=C_{X}[kiik]$, $i,k=1,2,3$, $X=(ee),(ll)$,
\begin{eqnarray}
&&C_{ld}[11ii],~ C_{lu}[11ii],~ C_{le}[11ii],~ C_{eu}[11ii],~C_{ed}[11ii],~C_{lq}^{(1)}[11ii],
\nonumber \\
&&\qquad 
~C_{lq}^{(3)}[11ii],~C_{qe}[ii11], C_{ee}[11ii],~C_{ll}[11ii],~\quad i=1,2,3\nonumber \\
&&C_{ee}[1jj1],~\quad  j=2,3\nonumber \\
&&C_{l q}^{(3)}[2233],~~C_{ll}[1331] 
 \, .
\end{eqnarray}
Finally, we note that there are 2 additional CP conserving
and 4 CP violating operators that contribute at NLO
\begin{eqnarray}
&{\text{CP~Conserving}}:& C_W\, C_\phi \nonumber \\
&{\text{CP~Violating}}: &C_{\tilde{W}}\, C_{\phi \tilde{W}}\, C_{\phi \tilde{B}}\,
C_{\phi \tilde{W}B}\, .
\end{eqnarray}
This yields a total of 66 CP conserving operators contributing at NLO.

At  NLO, we notice that in many cases the analytical contribution of different operators to the process is the same. Thus, we found it useful to write the results in terms of combinations of operators which contribute at NLO,
\begin{align}
\begin{split}
    {\hat{C}}^{2f}_1&= C_{\phi e }[22]+C_{\phi e }[33]+
    C_{\phi d }[11]+C_{\phi d}[22]+C_{\phi d }[33] \\ 
    &\hspace{10pt} -2 C_{\phi u}[11] -2 C_{\phi u}[22]
    -C_{\phi q}^{(1)}[11] -C_{\phi q}^{(1)}[22] +C_{\phi l}^{(1)}[33]
    \\
    {\hat{C}}^{2f}_2&= C_{\phi l}^{(3)}[33]+3C_{\phi q}^{(3)}[11] +3C_{\phi q}^{(3)}[22]
   \\
    \hat{C}^{4f}_1 &= C_{ld}[1111]+C_{ld}[1122]+C_{ld}[1133]-2\left(C_{lu}[1111]+C_{lu}[1122]\right)  \\ 
    &\hspace{10pt}-C_{lq}^{(1)}[1111]-C_{lq}^{(1)}[1122]+2C_{ll}[1133] \\ 
    &\hspace{10pt}+C_{le}[1122]+C_{le}[1133] -\frac{3M_W^2}{(M_Z^2-M_W^2)}\left(C_{lq}^{(3)}[1111]+C_{lq}^{(3)}[1122]\right) \, ,  \\ 
    \hat{C}^{4f}_{2} &=2(C_{ee}[1122]+C_{ee}[1133]) \\
    &\hspace{10pt}- C_{qe}[1111]-C_{qe}[2211]-2\left(C_{eu}[1111]+C_{eu}[1122]\right) \\ 
    &\hspace{10pt}+ C_{ed}[1111] + C_{ed}[1122]+C_{ed}[1133]+C_{le}[2211]+C_{le}[3311] \, ,  \\
    \hat{C}^{4f}_{3} &= 2(C_{ee}[1221]+C_{ee}[1331])\, .
\end{split}
 \end{align}

\begin{table}[h]
    \centering
    \begin{tabular}{|c|c|c|c|c|}
    \hline
Coefficient, $C_k$ & $\dfrac{\beta_k^{\left(0\right)}}{\Gamma^\text{SM}_\text{NLO}}\left(\dfrac{1~\text{TeV}^2}{\Lambda^2}\right)$ & $\dfrac{\beta_k^{\left(0\right)}+\beta_k^{\left(1\right)}}{\Gamma^\text{SM}_\text{NLO}}\left(\dfrac{1~\text{TeV}^2}{\Lambda^2}\right)$  & $\dfrac{\beta_k^{\left(1\right)}}{\Gamma^\text{SM}_\text{NLO}}\left(\dfrac{1~\text{TeV}^2}{\Lambda^2}\right)$ & $\dfrac{\beta_k^{\left(0\right)}+\beta_k^{\left(1\right)}}{\beta_{k}^{\left(0\right)}}$ \\
\hline
$C_{\phi B}$            &   -0.0851     &  -0.0749      & 0.010         & 0.88 \\
$C_{\phi D}$            &   0.0199      &  0.0152       & -0.0047       & 0.76 \\
$C_{\phi \square}$      &   0.119       &  0.126        & 0.0061        & 1.051  \\
$C_{\phi W}$            &   0.0268      &  0.0164       & -0.010        & 0.61 \\
$C_{\phi WB}$           &   -0.0510     &  -0.0474      & 0.0035        & 0.93 \\
$C_{\phi e}[ii]$       &   -0.00691    &  -0.00611     & 0.00080       & 0.88 \\
$C_{\phi l}^{(1)}[ii]$&    0.00859     &  0.00578      & -0.0028       & 0.67 \\
$C_{\phi l}^{(3)}[ii]$ &   -0.111      &  -0.117       & -0.0060       & 1.054  \\
$C_{\phi l}^{(3)}[jj]$ &   -0.119      &  -0.125       & -0.0056       & 1.047  \\
$C_{ll}[1221]$       &   0.119     &  0.123       &  0.0035        & 1.029  \\
\hline
    \end{tabular}
    \caption{Contributions from operators appearing at LO to the decay width $\Gamma(H\rightarrow \ell^+_i\ell^-_i Z)$ at LO and NLO to order $\mathcal{O}(1/\Lambda^2)$ for $\Lambda = 1$ TeV,  using the definition of eq.~\eqref{eq:smeftdef} and $\Gamma_\text{NLO}^\text{SM}=2.997\times 10^{-3}~\textrm{GeV}$. The flavor index $i$ corresponds to the produced lepton $(e,\mu)$, and $j$ is the lepton that is not produced, $(\mu,e$).}
    \label{tab:kfactors}
\end{table}

\begin{table}[]
    \centering
    \begin{tabular}{|c | c ||c|c|}
    \hline
    Coefficient & $\dfrac{\beta_k^{\left(1\right)}}{\Gamma^\text{SM}_{\text{NLO}}}\left(\dfrac{1~\text{TeV}^2}{\Lambda^2}\right)$ & Coefficient & $\dfrac{\beta_k^{\left(1\right)}}{\Gamma^\text{SM}_\text{NLO}}\left(\dfrac{1~\text{TeV}^2}{\Lambda^2}\right)$ \\
    \hline
$C_\phi$ & $-2.42\times 10^{-3}$ & $C_W$ & $6.58\times 10^{-4}$ \\
$C_{\phi l}^{(1)}[jj]$ & $-3.94\times 10^{-5}$ & $C_{\phi u}[33]$ & $-6.37\times 10^{-3} $ \\
$C_{\phi q}^{(1)}[33]$ & $5.66 \times 10^{-3}$ & $C_{\phi q}^{(3)}[33]$ & $-7.13 \times 10^{-3}$ \\
$C_{uW}[33]$ & $-2.73 \times 10^{-3}$ & $C_{uB}[33]$ & $2.86 \times 10^{-4}$ \\
$C_{u \phi}[33]$ & $8.09\times 10^{-4}$ & $\hat{C}_{1}^{2f}$ & $8.25 \times 10^{-5}$ \\
$\hat{C}_{2}^{2f}$ & $ -1.20 \times 10^{-4} $ & & \\
\hline
    \end{tabular}
    \caption{Contribution to the width $\Gamma(H\rightarrow \ell_i^+\ell_i^- Z)$ from SMEFT bosonic and two-fermion operators first appearing at NLO using the definitions of eq.~\eqref{eq:smeftdef}. We normalize to the NLO SM width and set $\Lambda = 1$ TeV. The flavor index $j$ corresponds to the lepton that is not produced.}
    \label{tab:virtonly_2f}
\end{table}

\begin{table}[]
    \centering
    \begin{tabular}{|c | c ||c|c|}
    \hline
    Coefficient & $\dfrac{\beta_k^{\left(1\right)}}{\Gamma^\text{SM}_{\text{NLO}}}\left(\dfrac{1~\text{TeV}^2}{\Lambda^2}\right)$ & Coefficient & $\dfrac{\beta_k^{\left(1\right)}}{\Gamma^\text{SM}_{\text{NLO}}}\left(\dfrac{1~\text{TeV}^2}{\Lambda^2}\right)$ \\
    \hline
$C_{eu}[ii33]$ & $-1.63\times 10^{-3}$ & $C_{qe}[33ii]$ & $1.70\times 10^{-3}$\\
$C_{lu}[ii33]$ & $2.02\times 10^{-3}$& $C_{lq}^{(1)}[ii33]$& $-2.11\times 10^{-3}$\\
$C_{lq}^{(3)}[ii33]$ & $2.31 \times 10^{-3}$ & $C_{lq}^{(3)}[jj33]$ & $3.25 \times 10^{-4}$\\
$C_{l e}[iiii]$ & $2.10\times 10^{-6}$ &$C_{ll}[iiii]$ & $3.01\times 10^{-7}$\\
$C_{ee}[iiii]$ & $-3.10\times 10^{-5}$ & $\hat{C}_{1}^{4f}$ & $1.07\times 10^{-5}$\\
$C_{ll}[1122]$ & $-2.61 \times 10^{-3}$& $C_{ll}[i33i]$ & $-2.12\times 10^{-5}$\\
$\hat{C}_{2}^{4f}$ & $-8.63 \times 10^{-6}$ & $\hat{C}_{3}^{4f}$ & $-6.85\times 10^{-6}$\\

\hline
    \end{tabular}
    \caption{Contribution to the width $\Gamma(H\rightarrow \ell_i^+\ell_i^- Z)$ from SMEFT four-fermion operators first appearing at NLO using the definitions of eq.~\eqref{eq:smeftdef}
    and $i=1,2$. We normalize to the NLO SM width and set $\Lambda = 1$ TeV. The flavor index $i$ corresponds to the produced lepton $(e,\mu)$, and $j$ is the lepton that is not produced, $(\mu,e$).}
    \label{tab:virtonly_4f}
\end{table}

\begin{table}[]
    \centering
    \begin{tabular}{|c | c ||c|c|}
    \hline
    Coefficient & $\dfrac{\widetilde{\beta}_k^{\left(1\right)}}{\widetilde{\Gamma}^\text{SM}_{\text{NLO}}}\left(\dfrac{1~\text{TeV}^2}{\Lambda^2}\right)$ & Coefficient & $\dfrac{\widetilde{\beta}_k^{\left(1\right)}}{\widetilde{\Gamma}^\text{SM}_{\text{NLO}}}\left(\dfrac{1~\text{TeV}^2}{\Lambda^2}\right)$ \\
    \hline

$C_\phi$                &  $-2.47 \times 10^{-3}$     & $C_W$                  &    $7.37\times 10^{-4}$      \\
$C_{\phi l}^{(1)}[jj]$  &  $-4.08\times 10^{-5}$     & $C_{\phi u}[33]$       &    $-6.50\times 10^{-3}$      \\
$C_{\phi q}^{(1)}[33]$  &  $5.79\times 10^{-3}$    & $C_{\phi q}^{(3)}[33]$ &       $-7.30\times 10^{-3}$   \\
$C_{uW}[33]$            &  $-2.40\times 10^{-3}$     & $C_{uB}[33]$           &    $2.13 \times 10^{-4}$      \\
$C_{u \phi}[33]$        &  $8.29\times 10^{-4}$     & $\hat{C}_{1}^{2f}$     &     $8.38\times 10^{-5}$     \\
$\hat{C}_{2}^{2f}$      &   $-1.23\times 10^{-4}$    & &                                 \\

\hline
    \end{tabular}
    \caption{Contribution to the width $\widetilde{\Gamma}(H\rightarrow e^+e^- Z)$ from SMEFT bosonic and two-fermion operators first appearing at NLO with a cut of $m_{ee}>12$ GeV using the definitions of eq.~\eqref{eq:smeftdef}. We normalize to the NLO SM width $\widetilde{\Gamma}^\text{SM}_\text{NLO}(H\rightarrow e^+e^-Z) = 2.582\times 10^{-3}$ MeV and set $\Lambda = 1$ TeV. The $H\rightarrow \mu^+\mu^-Z$ results can be obtained by rescaling these by $\widetilde{\Gamma}^\text{SM}_\text{NLO}(H\rightarrow \mu^+\mu^-Z) / \widetilde{\Gamma}^\text{SM}_\text{NLO}(H\rightarrow e^+e^-Z) = 0.987$. The flavor index $j$ corresponds to the lepton that is not produced.}
    \label{tab:virtonly_2f_qcut}
\end{table}

\begin{table}[]
    \centering
    \begin{tabular}{|c | c ||c|c|}
    \hline
    Coefficient & $\dfrac{\widetilde{\beta}_k^{\left(1\right)}}{\widetilde{\Gamma}^\text{SM}_{\text{NLO}}}\left(\dfrac{1~\text{TeV}^2}{\Lambda^2}\right)$ & Coefficient & $\dfrac{\widetilde{\beta}_k^{\left(1\right)}}{\widetilde{\Gamma}^\text{SM}_{\text{NLO}}}\left(\dfrac{1~\text{TeV}^2}{\Lambda^2}\right)$ \\
    \hline
$C_{eu}[ii33]$          &  $-1.64\times 10^{-3}$    & $C_{qe}[33ii]$        &  $1.72\times 10^{-3}$         \\
$C_{lu}[ii33]$          &  $2.03\times 10^{-3}$     & $C_{lq}^{(1)}[ii33]$  &  $-2.13\times 10^{-3}$         \\
$C_{lq}^{(3)}[ii33]$    &  $2.32\times10^{-3}$     & $C_{lq}^{(3)}[jj33]$  &  $3.32\times 10^{-4}$         \\
$C_{l e}[iiii]$         &  $2.38\times10^{-6}$    & $C_{ll}[iiii]$        &  $3.84\times 10^{-7}$         \\
$C_{ee}[iiii]$          &  $-3.51\times10^{-5}$   & $\hat{C}_{1}^{4f}$    &   $1.22\times 10^{-5}$        \\
$C_{ll}[1122]$     &  $-2.66\times 10^{-3}$   & $C_{ll}[i33i]$   &  $-2.40\times 10^{-5}$        \\
$\hat{C}_{2}^{4f}$      &  $-9.79\times 10^{-6}$     & $\hat{C}_{3}^{4f}$    & $-7.75\times 10^{-6}$          \\

\hline
    \end{tabular}
    \caption{Contribution to the width $\widetilde{\Gamma}(H\rightarrow e^+e^- Z)$ from SMEFT four-fermion operators first appearing at NLO with a cut of $m_{ee}>12$ GeV using the definitions of eq.~\eqref{eq:smeftdef}. We normalize to the SM width $\widetilde{\Gamma}^\text{SM}_\text{NLO}(H\rightarrow e^+e^-Z) = 2.582\times 10^{-3}$ MeV and set $\Lambda = 1$ TeV. The $H\rightarrow \mu^+\mu^-Z$ results can be obtained by rescaling these by $\widetilde{\Gamma}^\text{SM}_\text{NLO}(H\rightarrow \mu^+\mu^-Z) / \widetilde{\Gamma}^\text{SM}_\text{NLO}(H\rightarrow e^+e^-Z) = 0.987$. The flavor index $i$ corresponds to the produced lepton $(e,\mu)$, and $j$ is the lepton that is not produced, $(\mu,e$).}
    \label{tab:virtonly_4f_qcut}
\end{table}

\begin{table}[]
    \centering
    \begin{tabular}{|c|c|c|c|c|}
    \hline
Coefficient, $C_k$ & $\dfrac{\widetilde{\beta}_k^{\left(0\right)}}{\widetilde{\Gamma}^\text{SM}_\text{NLO}}\left(\dfrac{1~\text{TeV}^2}{\Lambda^2}\right)$ & $\dfrac{\widetilde{\beta}_k^{\left(0\right)}+\widetilde{\beta}_k^{\left(1\right)}}{\widetilde{\Gamma}^\text{SM}_\text{NLO}}\left(\dfrac{1~\text{TeV}^2}{\Lambda^2}\right)$  & $\dfrac{\widetilde{\beta}_k^{\left(1\right)}}{\widetilde{\Gamma}^\text{SM}_\text{NLO}}\left(\dfrac{1~\text{TeV}^2}{\Lambda^2}\right)$ & $\dfrac{\widetilde{\beta}_k^{\left(0\right)}+\widetilde{\beta}_k^{\left(1\right)}}{\widetilde{\beta}_{k}^{\left(0\right)}}$ \\
\hline
$C_{\phi B}$            &  -0.0830       & -0.0392       &   0.044       & 0.473   \\
$C_{\phi D}$            &  0.0203        & 0.0112        &   -0.0091     & 0.551   \\
$C_{\phi \square}$      &  0.122         & 0.126         &   0.0035      & 1.029   \\
$C_{\phi W}$            &  0.0172        & -0.0260       &   -0.043      & -1.52   \\
$C_{\phi WB}$           &  -0.0511       & -0.0271       &   0.0241      & 0.529   \\
$C_{\phi e}[11]$       &  -0.00789      & -0.00712      &   0.00077     & 0.903   \\
$C_{\phi l}^{(1)}[11]$ &  0.00980       & 0.00687       &   -0.0029     & 0.700   \\
$C_{\phi l}^{(3)}[11]$ &  -0.112        & -0.115        &   -0.0023     & 1.020   \\
$C_{\phi l}^{(3)}[22]$ &  -0.122        & -0.124        &   -0.0018     & 1.015   \\
$C_{ll}[1221]$       & 0.122        & 0.122        &   -0.00035   & 0.997   \\
\hline
    \end{tabular}
    \caption{Contributions from operators appearing at LO to the decay width $\widetilde{\Gamma}(H\rightarrow e^+e^- Z)$ at LO and NLO with a cut of $m_{ee}>12$ GeV using the definitions of eq.~\eqref{eq:smeftdef}.}
    \label{tab:kfactors_q12_ee}
\end{table}

\begin{table}[]
    \centering
    \begin{tabular}{|c|c|c|c|c|}
    \hline
Coefficient, $C_k$ & $\dfrac{\widetilde{\beta}_k^{\left(0\right)}}{\widetilde{\Gamma}^\text{SM}_\text{NLO}}\left(\dfrac{1~\text{TeV}^2}{\Lambda^2}\right)$ & $\dfrac{\widetilde{\beta}_k^{\left(0\right)}+\widetilde{\beta}_k^{\left(1\right)}}{\widetilde{\Gamma}^\text{SM}_\text{NLO}}\left(\dfrac{1~\text{TeV}^2}{\Lambda^2}\right)$  & $\dfrac{\widetilde{\beta}_k^{\left(1\right)}}{\widetilde{\Gamma}^\text{SM}_\text{NLO}}\left(\dfrac{1~\text{TeV}^2}{\Lambda^2}\right)$ & $\dfrac{\widetilde{\beta}_k^{\left(0\right)}+\widetilde{\beta}_k^{\left(1\right)}}{\widetilde{\beta}_{k}^{\left(0\right)}}$ \\
\hline
$C_{\phi B}$            &  -0.0819     &   -0.0399    &  0.042       & 0.488  \\
$C_{\phi D}$            &  0.0200      &   0.00994    &  -0.010      & 0.496  \\
$C_{\phi \square}$      &  0.120       &   0.125      &  0.0050      & 1.042  \\
$C_{\phi W}$            &  0.0169      &   -0.0251    &  -0.042      & -1.480  \\
$C_{\phi WB}$           &  -0.0505     &   -0.0304    &  0.020       & 0.601  \\
$C_{\phi e}[22]$       &  -0.00779    &   -0.00709   &  0.00070     & 0.910  \\
$C_{\phi l}^{(1)}[22]$ &  0.00968     &   0.00685    &  -0.0028     & 0.708  \\
$C_{\phi l}^{(3)}[22]$ &  -0.111      &   -0.115     &  -0.0045     & 1.041  \\
$C_{\phi l}^{(3)}[11]$ &  -0.120      &   -0.125     &  -0.0041     & 1.034  \\
$C_{ll}[1221]$      & 0.120      &   0.122     &  0.0020      & 1.017  \\
\hline
    \end{tabular}
    \caption{Contributions from operators appearing at LO to the decay width $\widetilde{\Gamma}(H\rightarrow \mu^+\mu^- Z)$ at LO and NLO with a cut of $m_{\mu\mu}>12$ GeV  using the definitions of eq.~\eqref{eq:smeftdef}.}
    \label{tab:kfactors_q12_mumu}
\end{table}

\end{document}